\documentstyle[aps,multicol,amsfonts,amssymb,amsmath,tcvn,graphicx]{revtex}

\makeatletter
\def\thepage{\@arabic\c@page}
\def\@pnumwidth{2em}
\makeatother

\begin{document}


\title{\Huge Ph­¬ng tr×nh Nh©n qu¶}
\author{§ç Minh ChÝ}
\address{Hanoi, Vietnam.}
\date{1979}
\maketitle



\makeatletter
\global\@specialpagefalse
\def\@oddhead{§ç Minh ChÝ\hfill Ph­¬ng tr×nh Nh©n qu¶}
\let\@evenhead\@oddhead
\def\@oddfoot{\reset@font\rm\hfill \thepage\hfill
\ifnum\c@page=1
  \llap{\protect\copyright{} 1979 §. M. ChÝ}%
\fi
} \let\@evenfoot\@oddfoot
\makeatother

\vspace{1cm}

\begin{multicols}{2}

NÕu thÕ giíi lµ {\it thèng nhÊt} th× nã thèng nhÊt trong mèi quan
hÖ {\it nh©n qu¶} vµ sù thèng nhÊt chØ cã thÓ ®­îc biÓu hiÖn trong
ý nghÜa Êy mµ th«i.

Theo tinh thÇn ®ã, sù ngÉu nhiªn, nÕu nh­ thËt sù cã c¸i g× ®ã lµ
ngÉu nhiªn, còng chØ lµ s¶n phÈm cña sù tÊt yÕu.

Bëi thÕ giíi lµ thèng nhÊt trong mèi quan hÖ nh©n qu¶, kh«ng mét
c¸i g× cña thÕ giíi n»m ngoµi mèi quan hÖ Êy, nªn ta cã thÓ chia
toµn bé thÕ giíi thµnh hai tËp hîp: tËp hîp $A$ bao gåm tÊt c¶
nh÷ng g× ®­îc coi lµ nguyªn nh©n, tËp hîp $B$ bao gåm tÊt c¶ nh÷ng
g× lµ hÖ qu¶.

Chóng ta h·y lo¹i bá khái c¶ hai tËp hîp tÊt c¶ c¸c phÇn tö gièng
nhau. Nh­ thÕ sÏ cã 4 kh¶ n¨ng sau:

\begin{enumerate}
\item C¶ hai tËp hîp $A$, $B$ ®Òu trë thµnh c¸c tËp hîp rçng, tøc
lµ kh«ng cã nguyªn nh©n thuÇn tuý còng nh­ hÖ qu¶ thuÇn tuý. Nãi
c¸ch kh¸c, thÕ giíi kh«ng cã më ®Çu, còng kh«ng cã kÕt côc cuèi
cïng.

\item $A$ kh«ng rçng, $B$ rçng. Nh­ vËy tån t¹i nguyªn nh©n thuÇn
tuý - thÕ giíi cã b¾t ®Çu nh­ng kh«ng cã kÕt côc cuèi cïng.

\item $A$ rçng, $B$ kh«ng rçng. Kh«ng cã nguyªn nh©n thuÇn tuý
nh­ng cã hÖ qu¶ thuÇn tuý. ThÕ giíi kh«ng cã më ®Çu nh­ng cã kÕt
côc cuèi cïng.

\item C¶ $A$, c¶ $B$ ®Òu kh«ng rçng. ThÕ giíi cã më ®Çu vµ cã kÕt
côc cuèi cïng.
\end{enumerate}

ChØ mét trong 4 kh¶ n¨ng trªn lµ ®óng víi hiÖn thùc. §ã lµ kh¶
n¨ng nµo vµ ®iÒu ®ã phô thuéc vµo c¸i g×?

Mét ®éng lùc ®Çy bÝ Èn lu«n lu«n th«i thóc con ng­êi t×m kiÕm
nguyªn nh©n cho mäi hiÖn t­îng vµ sù vËt. Dßng t­ t­ëng duy t©m
(kh¸ch quan vµ chñ quan) cho r»ng ý niÖm tuyÖt ®èi, tinh thÇn tèi
cao hay ®Êng s¸ng t¹o, th­îng ®Õ,... lµ nguyªn nh©n tèi cao, lµ
nguyªn nh©n cña tÊt c¶. Dßng t­ t­ëng duy vËt l¹i cho r»ng vËt
chÊt míi lµ nguån gèc cña tÊt c¶, lµ c¸i ph¶i cã tr­íc nhÊt.

HiÖn tr¹ng ®ã lµ {\it m©u thuÉn}!

NÕu qu¶ thùc tån t¹i mét nguyªn nh©n tèi cao, th× ®ã lµ {\it sù
kh¸c biÖt}!

Qu¶ vËy, nÕu kh«ng tån t¹i sù kh¸c biÖt, th× sÏ kh«ng tån t¹i bÊt
cø c¸i g×, kÓ c¶ dßng t­ t­ëng duy t©m víi nh÷ng ý niÖm vµ tinh
thÇn cña nã, kÓ c¶ dßng t­ t­ëng duy vËt víi c¸c c¬ së vËt chÊt
cña nã. Tãm l¹i, nÕu kh«ng cã sù kh¸c biÖt th× kh«ng cã thÕ giíi
nµy.

ThÕ nh­ng, nÕu kh¸c biÖt lµ nguyªn nh©n tèi cao, tøc lµ nguyªn
nh©n cña mäi nguyªn nh©n th× nã ph¶i lµ nguyªn nh©n cña chÝnh nã
n÷a, hay nãi kh¸c h¬n, nã lµ hÖ qu¶ cña chÝnh nã.

Chóng ta ®· thõa nhËn sù tån t¹i cña kh¸c biÖt, ®iÒu ®ã cã nghÜa
lµ chóng ta ngÇm thõa nhËn {\it tÝnh b¶o toµn t­¬ng ®èi} cña nã:
Qu¶ vËy, nÕu lóc nµy b¹n lµ nhµ duy vËt th× b¹n kh«ng thÓ còng lµ
kÎ duy t©m ®­îc n÷a. Qu¶ thÞ, khi chøa c« TÊm ë bªn trong, kh«ng
ph¶i lµ qu¶ thÞ theo ®óng nghÜa cña nã. C¸i bµn kia, chõng nµo nã
cßn lµ c¸i bµn, th× nã kh«ng thÓ lµ trang giÊy mµ b¹n ®ang ®äc
®­îc!.

\begin{center}
{\bf C©u chuyÖn x¶y ra trong h×nh häc}
\end{center}

Chóng ta trë l¹i mét c©u chuyÖn ®· cò: cuéc tranh c·i xung quanh
hÖ tiªn ®Ò h×nh häc Euclid.

VÉn do chÝnh c¸i ®éng lùc bÝ Èn duy nhÊt Êy chi phèi mµ ng­êi ta
lu«n khao kh¸t t×m kiÕm c¸i "nguyªn nh©n tèi cao". {\VnTimeH ë}
®©y môc ®Ých ®ã khiªm tèn h¬n, chØ giíi h¹n trong ph¹m vi h×nh
häc, vµ ng­êi ®Çu tiªn thùc hiÖn ®­îc ®iÒu ®ã lµ Euclid.

Bèn tiªn ®Ò ®Çu cña «ng ®­îc dÔ dµng chÊp nhËn v× chóng râ rµng vµ
hiÓn nhiªn, nh­ng tiªn ®Ò thø n¨m, cßn gäi lµ ®Þnh ®Ò Euclid, ®·
khiÕn ng­êi ta nghi ngê vÒ b¶n chÊt tiªn ®Ò cña nã: "®Þnh ®Ò nµy
phøc t¹p vµ kÐm hiÓn nhiªn".

Trong suèt h¬n 20 thÕ kû c¸c nhµ to¸n häc mäi thêi ®¹i ®· g¾ng søc
chøng minh nã chØ lµ mét ®Þnh lý. Nh­ng mäi cè g¾ng ®Òu v« Ých. Vµ
kÕt qu¶ lµ dÉn tíi sù ra ®êi mét h×nh häc míi: h×nh häc phi
Euclid.

B»ng c¸ch ®ã, ng­êi ta ®· kh¼ng ®Þnh r»ng, vÊn ®Ò nh­ thÕ lµ ®·
®­îc gi¶i quyÕt: ®Þnh ®Ò Euclid ®óng lµ mét tiªn ®Ò, v× r»ng ®iÒu
gi¶ thiÕt ng­îc l¹i ®· dÉn tíi h×nh häc phi Euclid kh«ng cã m©u
thuÉn néi t¹i.

Nh­ng... kÕt luËn nh­ vËy cã dÔ d·i qu¸ kh«ng?

Khi ng­êi ta hoan hØ v× h×nh nh­ mäi viÖc ®· ®©u vµo ®Êy vµ môc
®Ých ®Ò ra ®· ®­îc thùc hiÖn: gi¶m ®Õn møc tèi thiÓu sè l­îng tiªn
®Ò cña h×nh häc vµ lµm trong s¸ng chóng, th× trí trªu thay, mét
tiªn ®Ò míi l¹i ®­îc lÐn ®­a thªm vµo mét c¸ch ®­êng hoµng: tiªn
®Ò Lobachevski. Tiªn ®Ò nµy vµ tiªn ®Ò thø n¨m Euclid lo¹i trõ lÉn
nhau!

Kh«ng mét ai nhËn thøc râ rµng vµ s©u s¾c m©u thuÉn nµy nghÜa lµ
thÕ nµo. Nh­ng m©u thuÉn vÉn lµ m©u thuÉn: nã g©y ra bao cuéc
tranh c·i vµ ph¶n b¸c kÞch liÖt, thËm chÝ c¶ sù h»n häc.

Sau nµy, khi Bentrami ®· chøng minh ®­îc sù ®óng ®¾n cña h×nh häc
Lobachevski trªn mÆt gi¶ cÇu, sù ph¶n b¸c cã dÞu xuèng.

NÕu ngay tõ ®Çu, c¸c nhµ h×nh häc phi Euclid khi b¾t tay vµo x©y
dùng h×nh häc cña m×nh, tuyªn bè râ víi ®éc gi¶ r»ng: ®èi t­îng
cña h×nh häc míi kh«ng ph¶i lµ mÆt ph¼ng Euclid mµ lµ mÆt gi¶ cÇu,
kh«ng ph¶i lµ ®­êng th¼ng Euclid mµ lµ ®­êng th¼ng cña mÆt gi¶ cÇu
th× cã lÏ ®· ch¼ng cã ai th¾c m¾c vµ ph¶n ®èi g× c¶!

ThËt ®¸ng tiÕc! hay kh«ng ®¸ng tiÕc lµ ®· kh«ng x¶y ra nh­ vËy?

Nh­ng ®iÒu ®¸ng tiÕc thËt sù lµ: toµn bé vÊn ®Ò l¹i kh«ng n»m
trong c¸i ®· ®­îc ®­a ra vµ gi¶i quyÕt ngoµi s©n khÊu mµ ë hËu qu¶
cña nã trong hËu tr­êng. Bëi v×, cho dï h×nh häc phi Euclid cã
tuyÖt ®èi ®óng ë ®©u ch¨ng n÷a, th× ®iÒu ®ã cã nghÜa lµ: còng cïng
nh÷ng ®èi t­îng Êy cña h×nh häc - mÆt ph¼ng vµ ®­êng th¼ng Euclid
- vËy mµ gi÷a chóng cã thÓ tån t¹i hai kiÓu quan hÖ lo¹i trõ nhau!
®­îc diÔn ®¹t trong c¸c tiªn ®Ò Euclide vµ tiªn ®Ò Lobachevski.

Cã thÓ viÖn ®Õn c¸c lý lÏ nµy hoÆc kh¸c ®Ó Ðp m×nh chÊp nhËn ®iÒu
khã chÞu nµy, nh­ng ®ã sÏ kh«ng ph¶i lµ sù trung thùc. {\VnTimeH
ë} ®©y, tÝnh ®¬n trÞ nh©n qu¶ ®· bÞ ph¸ vì; ë ®©y tÝnh b¶o toµn
t­¬ng ®èi cña sù kh¸c biÖt ®· bÞ lÉn lén tr¾ng ®en, cã nguy c¬
trang giÊy còng lµ c¸i bµn, vµ c¸i bµn lµ trang giÊy.

Ph­¬ng ph¸p tiªn ®Ò ®­îc sö dông réng r·i trong to¸n häc râ rµng
®· mang l¹i nhiÒu tiÖn lîi, nh­ng ph­¬ng ph¸p ®ã chØ tèt khi tÝnh
®¬n trÞ nh©n qu¶ ®­îc b¶o ®¶m, khi chóng ta lu«n lu«n chó ý ®Ó
kh«ng t­íc ®o¹t mÊt ë c¸c ®èi t­îng kh¶o s¸t ý nghÜa thùc, vËt lý
cña chóng. NÕu ®­a ra c¸c kiÓu quan hÖ cña c¸c ®èi t­îng d­íi d¹ng
c¸c tiªn ®Ò mµ bÊt kÓ ®Õn c¸c ®èi t­îng - chñ nh©n thùc sù cña c¸c
mèi quan hÖ ®ã th× rÊt cã thÓ, ë mét chç kh«ng ngê nhÊt, tÝnh ®¬n
trÞ nh©n qu¶ bÞ ph¸ vì vµ m©u thuÉn n¶y sinh.

Bëi v× c¸i mµ chóng ta thèng nhÊt víi nhau lµ: c¸c ®èi t­îng lµ
c¸i cã tr­íc, c¸c mèi quan hÖ cña c¸c ®èi t­îng lµ hÖ qu¶ tÊt yÕu
do sù tån t¹i ®ång thêi cña c¸c ®èi t­îng g©y ra, chø kh«ng ph¶i
ng­îc l¹i.

NÕu ta cã mét tËp hîp c¸c ®èi t­îng vµ ta muèn t×m kiÕm tÊt c¶ c¸c
mèi quan hÖ cã thÓ gi÷a chóng b»ng ph­¬ng ph¸p lËp luËn logic th×
cã lÏ tr­íc hÕt vµ Ýt ra ta ph¶i biÕt mèi quan hÖ néi t¹i cña tõng
lo¹i ®èi t­îng.

C¸c mèi quan hÖ néi t¹i quyÕt ®Þnh b¶n chÊt cña ®èi t­îng, ®Õn
l­ît m×nh b¶n chÊt cña c¸c ®èi t­îng quyÕt ®Þnh c¸c mèi quan hÖ cã
thÓ cã gi÷a chóng, vµ tÊt nhiªn gi÷a chóng kh«ng thÓ ®ång thêi tån
t¹i c¸c mèi quan hÖ lo¹i trõ nhau.

Mèi quan hÖ néi t¹i, theo c¸ch nãi cña c¸c triÕt gia, lµ tÝnh tù
th©n cña sù vËt. Khoa häc ngµy nay chÝnh lµ ®ang t×m kiÕm c¸i tÝnh
tù th©n ®ã cña sù vËt trªn c¶ hai h­íng, réng h¬n vµ c¬ b¶n h¬n.

Trë l¹i m¹ch cña c©u chuyÖn, nh­ chóng ta ®· thÊy, còng chÝnh
nh÷ng ®èi t­îng Êy cña h×nh häc Euclid mµ l¹i cã hai kiÓu quan hÖ
lo¹i trõ nhau, ®iÒu ®ã ph¶i ®­îc hiÓu nh­ thÕ nµo?

ChØ cã thÓ lµ hÖ tiªn ®Ò Euclid lµ ch­a ®Çy ®ñ theo nghÜa: sù hiÓu
vÒ c¸c ®èi t­îng cña h×nh häc nµy lµ ch­a hoµn h¶o. B¶n th©n
Euclid còng ®· tõng ®­a ra c¸c ®Þnh nghÜa vÒ c¸c ®èi t­îng h×nh
häc cña «ng, nh­ng nh­ c¸c nhµ to¸n häc hiÖn ®¹i ®· phª ph¸n lµ:
"lóng tóng", "nÆng tÝnh trùc gi¸c". Theo hä, c¸c ®èi t­îng xuÊt
ph¸t cña h×nh häc lµ kh«ng ®Þnh nghÜa ®­îc vµ chØ ®¬n thuÇn gäi
chóng lµ ®iÓm, ®­êng, mÆt v.v. v× lý do t«n träng lÞch sö mµ th«i.

Nh­ng, c¸c ®èi t­îng h×nh häc cßn cã c¸c tªn gäi kh¸c n÷a: c¸c
kh«ng gian "kh«ng chiÒu", "mét chiÒu", "hai chiÒu", vµ "ba chiÒu".
(Kh«ng gian "kh«ng chiÒu" tøc ®iÓm, lµ do t¸c gi¶ ®­a vµo cho
"trän bé".)

Ta cã thÓ hái r»ng, c¸c ®èi t­îng nµy cã thÓ tù tån t¹i ®éc lËp
víi nhau ®­îc kh«ng, nÕu cã th× v× sao chóng cã thÓ quan hÖ víi
nhau ®­îc?

LÇn theo m¹ch logic cña sù viÖc, ta thÊy r»ng c¸c kh¸i niÖm vÒ
nh÷ng ®èi t­îng nµy n¶y sinh ra tõ kinh nghiÖm thu ®­îc qua ho¹t
®éng thùc tiÔn cña con ng­êi trong Tù nhiªn chø kh«ng ph¶i do bÈm
sinh, tù chóng cã s½n trong ®Çu cña chóng ta (Bëi vËy, ta kh«ng
nªn t¸ch rêi chóng khái trùc gi¸c, kh«ng nªn t­íc ®o¹t kh¶ n¨ng
h×nh dung ra chóng; ®iÒu ®ã phi lý biÕt bao!)

Nh×n nhËn ë møc ®é s©u h¬n, cã thÓ thÊy r»ng, kh«ng ph¶i tÊt c¶
trong sè c¸c ®èi t­îng h×nh häc ®Òu cã thÓ tån t¹i ®éc lËp, mµ bÊt
kú mét kh«ng gian $n$ chiÒu nµo còng lµ giao cña hai kh«ng gian
kh¸c cã chiÒu lín h¬n mét ®¬n vÞ $n+1$.

VËy lµ, h×nh nh­ chóng ta cã ®Þnh nghÜa: ®iÓm lµ giao cña hai
®­êng, ®­êng lµ giao cña hai mÆt, mÆt lµ giao cña hai khèi, cßn
khèi ... lµ giao cña nh÷ng kh«ng gian nµo n÷a?

Tuy nhiªn, trong h×nh häc, nhê kh¶ n¨ng h×nh dung cña n·o, chóng
mÆc nhiªn trë thµnh c¸c ®èi t­îng ®éc lËp, vµ ®Ó gi¶n tiÖn, ta sÏ
gäi chóng lµ c¸c thùc thÓ kh«ng gian.

C¸c ®èi t­îng ®¬n gi¶n nhÊt cña h×nh häc lµ c¸c thùc thÓ ®ång
nhÊt. §ã lµ c¸c thùc thÓ, mµ nãi mét c¸ch gi¶n ®¬n, khi ta dÞch
chuyÓn trªn chóng theo tÊt c¶ c¸c bËc tù do cã thÓ cã cña chóng,
ta kh«ng thÓ ph¸t hiÖn ®­îc bÊt kú sù kh¸c biÖt néi t¹i nµo trong
chóng.

C¸c ®èi t­îng cña h×nh häc Euclid lµ mét bé phËn cña tËp hîp c¸c
thùc thÓ ®ång nhÊt. NÕu ta x©y dùng hÖ tiªn ®Ò chØ riªng cho bé
phËn nµy th× râ rµng hÖ ®ã sÏ kh«ng tæng qu¸t.

HÖ tiªn ®Ò dïng cho c¸c ®èi t­îng kh«ng gian ®ång nhÊt chÝnh lµ hÖ
tiªn ®Ò dïng cho mÆt cÇu. H×nh häc Euclid chØ lµ tr­êng hîp giíi
h¹n cña h×nh häc tæng qu¸t nµy.

§èi víi mÆt cÇu, tøc lµ mÆt ®ång nhÊt ë d¹ng tæng qu¸t, tån t¹i
®Þnh ®Ò sau: hai ®­êng th¼ng bÊt kú kh«ng trïng nhau (®­êng th¼ng
lµ ®­êng ®ång nhÊt chia mÆt chøa nã thµnh hai phÇn b»ng nhau) bao
giê còng c¾t nhau t¹i hai ®iÓm vµ hai ®iÓm ®ã chia ®«i mçi ®­êng.

Cã thÓ ph¸t biÓu kh¸c h¬n: hai ®iÓm bÊt kú trªn mét mÆt ®ång nhÊt
chØ thuéc vÒ mét ®­êng th¼ng duy nhÊt còng trªn mÆt ®ã nÕu chóng
kh«ng chia ®­êng nµy thµnh hai phÇn b»ng nhau.

{\VnTimeH ¸}p dông ®Þnh ®Ò nµy cho mÆt ph¼ng Euclid, lµ tr­êng hîp
giíi h¹n, ta thÊy ngay r»ng ®©y chÝnh lµ néi dung cña tiªn ®Ò thø
nhÊt Euclid: qua hai ®iÓm chØ cã thÓ kÎ ®­îc mét ®­êng th¼ng duy
nhÊt mµ th«i. Qu¶ vËy, hai ®iÓm bÊt kú trong ph¹m vi kh¶o s¸t ®­îc
cña mÆt ph¼ng Euclid chØ thuéc vÒ mét ®­êng th¼ng duy nhÊt v×
chóng kh«ng chia ®­êng ®ã thµnh hai phÇn b»ng nhau.

Nh­ vËy, cã thÓ nãi r»ng, c¸ch ph¸t biÓu ®Þnh ®Ò 5 Euclid lµ kh«ng
chÝnh x¸c ngay tõ ®Çu, bëi v× bÊt kú hai ®­êng th¼ng nµo cña mÆt
®ång nhÊt ®· cho lu«n lu«n c¾t nhau t¹i hai ®iÓm vµ chia ®«i nhau.
Trªn mÆt ph¼ng Euclid ta chØ thÊy hoÆc mét giao ®iÓm cña chóng cßn
®iÓm kia ë ngoµi v« cïng; hoÆc lµ ta kh«ng thÊy ®iÓm nµo c¶ -
chóng ®Òu n»m ë ngoµi v« cùc. Trong tr­êng hîp ®ã hai ®­êng th¼ng
®­îc gäi lµ song song (biÓu kiÕn) víi nhau.

C¸c c¸ch ph¸t biÓu t­¬ng ®­¬ng cña ®Þnh ®Ò 5, sau khi ®­îc chÝnh
x¸c ho¸ theo tinh thÇn cña nhËn xÐt trªn, ®Òu cã thÓ ®­îc chøng
minh nh­ mét ®Þnh lý.

Mét tÝnh chÊt rÊt quan träng cña c¸c thÓ kh«ng gian lµ: mét thùc
thÓ kh«ng gian bÊt kú chØ cã thÓ bÞ chøa trong mét thùc thÓ kh«ng
gian kh¸c ®ång chiÒu, ®ång ®é cong hoÆc cã chiÒu lín h¬n nh­ng ®é
cong kh«ng lín h¬n.

§iÒu nµy h×nh nh­ ®· qu¸ hiÓn nhiªn: hai ®­êng trßn cã ®é cong
kh¸c nhau th× kh«ng thÓ chøa trong nhau ®­îc; mét mÆt trßn cã ®é
cong lín h¬n kh«ng thÓ chøa mét ®­êng trßn cã ®é cong nhá h¬n...

T­¬ng tù nh­ vËy, hai kh«ng gian (khèi) cã ®é cong kh¸c nhau th×
kh«ng thÓ chøa trong nhau ®­îc. (Ta h·y trë l¹i vÝ dô vÒ qu¶ thÞ
vµ c« TÊm, vÒ c¸i bµn vµ trang giÊy...). §é cong ë ®©y t­¬ng øng
víi mét ®¹i l­îng nµo ®ã ®Æc tr­ng cho mèi quan hÖ néi t¹i cña ®èi
t­îng kh¶o s¸t.

\begin{center}
{\bf M©u thuÉn, sinh ra trªn c¬ së cña sù kh¸c biÖt,} \\ {\bf lµ
nguån ®éng lùc cña tÊt c¶}
\end{center}

VÒ thùc chÊt, Tù nhiªn lµ mét tËp hîp cña c¸c {\it kh¼ng ®Þnh} vµ
{\it phñ ®Þnh}.

VËy Tù nhiªn kh¼ng ®Þnh nh÷ng g× vµ phñ ®Þnh nh÷ng g×?

Nh÷ng bÝ mËt ®ã ngµy cµng ®­îc khoa häc kh¸m ph¸, ph¸t hiÖn vµ
trong cuéc t×m kiÕm ®ã, nÕu kh«ng kÓ ®Õn nguån gèc ®éng lùc cña
nã, lËp luËn logic ®ãng vai trß to lín.

Nh­ng c¸i mµ ta gäi lµ logic ph¶i ch¨ng kh«ng ph¶i lµ mét chuçi
nh÷ng kh¼ng ®Þnh vµ phñ ®Þnh ë c¸c cÊp bËc vµ tæ hîp kh¸c nhau?

VËy th× khi nµo c¸i nµy ®­îc kh¼ng ®Þnh, cßn c¸i kia th× kh«ng?

Bëi v× t­ duy còng chØ lµ mét hiÖn t­îng cña Tù nhiªn, nªn quy
luËt kh¼ng ®Þnh vµ phñ ®Þnh cña t­ duy còng lµ quy luËt kh¼ng ®Þnh
vµ phñ ®Þnh cña Tù nhiªn. Nãi c¸ch kh¸c, quy luËt kh¼ng ®Þnh vµ
phñ ®Þnh cña Tù nhiªn ®· ph¶n ¸nh vµ thÓ hiÖn qua chÝnh quy luËt
kh¼ng ®Þnh vµ phñ ®Þnh cña t­ duy.

Quy luËt ®ã lµ: {\it c¸i g× kh«ng cã m©u thuÉn néi t¹i th× ®­îc
kh¼ng ®Þnh, c¸i g× chøa m©u thuÉn néi t¹i th× bÞ phñ ®Þnh}.

Sù kh¼ng ®Þnh (nÕu nh×n vÒ phÝa tr­íc cña qu¸ tr×nh) hay sù phñ
®Þnh (nÕu nh×n vÒ phÝa sau cña tiÕn tr×nh) ®Òu cã ®Ých cuèi cïng
lµ ®¹t tíi vµ kÕt thóc ë mét kh¼ng ®Þnh míi.

Chóng ta h·y lÊy mét líp nh÷ng kh¸i niÖm rÊt gÇn nhau lµ: cã, tån
t¹i, b¶o toµn, kh¼ng ®Þnh.

§èi lËp l¹i víi chóng lµ líp kh¸i niÖm phñ ®Þnh cña chóng: kh«ng,
kh«ng tån t¹i, kh«ng b¶o toµn, phñ ®Þnh.

Chóng n»m trong sè nh÷ng kh¸i niÖm tæng qu¸t nhÊt, c¬ b¶n nhÊt,
bëi trong bÊt kú hiÖn t­îng nµo cña Tù nhiªn: c¶m gi¸c, suy nghÜ,
vËn ®éng, biÕn ®æi v.v. ®Òu lu«n lu«n cã sù biÓu hiÖn cña chóng.

Nh­ng ho¸ ra lµ søc m¹nh cña hai líp kh¸i niÖm nµy kh«ng t­¬ng
®­¬ng nhau (vµ ®ã lµ mét ®iÒu thËt may m¾n).

Chóng ta h·y thiÕt lËp kh¼ng ®Þnh sau gäi lµ {\it kh¼ng ®Þnh} $A$:

"Tån t¹i tÊt c¶, cã tÊt c¶, b¶o toµn tÊt c¶, kh¼ng ®Þnh tÊt c¶".

Cßn {\it kh¼ng ®Þnh} $B$ cã néi dung ng­îc l¹i:

"Kh«ng cã c¸i g× hÕt, kh«ng tån t¹i bÊt cø c¸i g×, kh«ng b¶o toµn
c¸i g× c¶, phñ ®Þnh tÊt c¶".

Kh¼ng ®Þnh {\it kh¼ng ®Þnh} $A$, tøc lµ phñ ®Þnh {\it kh¼ng ®Þnh}
$B$. Vµ ng­îc l¹i.

{\it Kh¼ng ®Þnh} $B$ nãi r»ng:

- Kh«ng cã c¸i g× hÕt, tøc lµ kh«ng cã chÝnh {\it kh¼ng ®Þnh} $B$.

- Kh«ng tån t¹i c¸i g×, tøc lµ kh«ng tån t¹i b¶n th©n {\it kh¼ng
®Þnh} $B$.

- Kh«ng b¶o toµn c¸i g×, vËy lµ chÝnh {\it kh¼ng ®Þnh} $B$ còng
kh«ng ®­îc b¶o toµn.

- Phñ ®Þnh tÊt c¶, vËy lµ phñ ®Þnh chÝnh {\it kh¼ng ®Þnh} $B$.

Tãm l¹i, {\it kh¼ng ®Þnh} $B$ chøa m©u thuÉn néi t¹i. Nã tù phñ
®Þnh chÝnh m×nh. Tù phñ ®Þnh m×nh, {\it kh¼ng ®Þnh} $B$ mÆc nhiªn
kh¼ng ®Þnh {\it kh¼ng ®Þnh} $A$, vµ ®iÒu ®ã cã nghÜa lµ: kh«ng tån
t¹i h­ v« hay sù trèng rçng tuyÖt ®èi, vµ chÝnh bëi lÏ ®ã mµ thÕ
giíi ®· ®­îc sinh ra!

Cßn {\it kh¼ng ®Þnh} $A$ kh¼ng ®Þnh tÊt c¶, kÓ c¶ chÝnh nã lÉn
{\it kh¼ng ®Þnh} $B$, nh­ng {\it kh¼ng ®Þnh} $B$ tù phñ ®Þnh chÝnh
m×nh, nªn {\it kh¼ng ®Þnh} $A$ kh«ng chøa m©u thuÉn néi t¹i.

Nh­ vËy, trong khu«n khæ cña {\it kh¼ng ®Þnh} $A$ nh÷ng g× kh«ng
tù phñ ®Þnh th× ®­îc kh¼ng ®Þnh.

\begin{center}
{\bf C¸i g× c¬ b¶n h¬n}
\end{center}

Bèn kh¸i niÖm rÊt quan träng cña tri thøc lµ: {\it thêi gian},
{\it kh«ng gian}, {\it vËt chÊt} vµ {\it vËn ®éng}. Chóng kh¸c
biÖt nhau, nh­ng ph¶i ch¨ng chóng b×nh ®¼ng víi nhau vµ cã thÓ tån
t¹i ®éc lËp víi nhau?

Ta h·y b¾t ®Çu tõ thêi gian. Nã lµ mét thùc thÓ ch¨ng? nã cã thÓ
tån t¹i ®éc lËp, t¸ch rêi khái kh«ng gian, vËt chÊt vµ vËn ®éng
ch¨ng?. HiÓn nhiªn lµ kh«ng. ChØ cÇn c¸ch ly thêi gian khái vËn
®éng lµ kh¸i niÖm vÒ nã sÏ mÊt mäi ý nghÜa, thêi gian sÏ chÕt.
Kh¸i niÖm vËn ®éng cã tÝnh ®éc lËp cao h¬n so víi kh¸i niÖm thêi
gian.

VËy th× thêi gian ch­a ph¶i lµ c¸i ®Çu tiªn. Nã kh«ng tù tån t¹i
®­îc nªn nã chØ cã thÓ lµ hÖ qu¶ cña nh÷ng c¸i cßn l¹i.

VËn ®éng còng kh«ng ph¶i lµ c¸i c¬ b¶n ®Çu tiªn. Nã kh«ng thÓ tù
tån t¹i t¸ch rêi khái vËt chÊt vµ kh«ng gian. Thùc chÊt, vËn ®éng
chØ lµ mét biÓu hiÖn cña mèi quan hÖ gi÷a vËt chÊt vµ kh«ng gian.

VËy th× gi÷a hai c¸i cßn l¹i, vËt chÊt vµ kh«ng gian, c¸i nµo c¬
b¶n h¬n, c¸i nµo cã tr­íc, hay chóng b×nh ®¼ng víi nhau vµ cïng do
mét c¸i g× ®ã c¬ b¶n h¬n sinh ra? Cã lÏ ®Æt vÊn ®Ò nh­ vËy lµ
thõa, bëi v× còng nh­ thêi gian vµ vËn ®éng, vËt chÊt kh«ng thÓ
t¸ch rêi vµ tån t¹i ngoµi kh«ng gian. H·y lÊy mét biÓu hiÖn cô thÓ
cña vËt chÊt, ch¼ng h¹n, trang giÊy nµy. Nã tån t¹i kh«ng ph¶i chØ
nhê b¶n th©n nã, mµ cßn do sù tån t¹i ®ång thêi cña kh«ng gian bao
quanh nã (hay chøa nã), lµm cho trang giÊy vÉn cßn lµ trang giÊy.

Râ rµng lµ vËt chÊt còng thuéc vÒ ph¹m trï kh«ng gian vµ nã cã thÓ
lµ c¸i g× kh¸c h¬n n÷a nÕu kh«ng ph¶i lµ chÝnh kh«ng gian cã mèi
quan hÖ néi t¹i kh¸c víi c¸i kh«ng gian th«ng th­êng mµ ta vÉn
hiÓu?!

Nh­ng khi ®ã, theo tÝnh chÊt cña c¸c thÓ kh«ng gian ®· nãi ë trªn,
®iÒu nµy lµ m©u thuÉn: hai kh«ng gian ®ång chiÒu cã ®é cong (tøc
mèi quan hÖ néi t¹i) kh¸c nhau th× kh«ng thÓ chøa trong nhau ®­îc!

VËy th× hoÆc lµ chóng ta ®· sai: hiÓn nhiªn lµ cã thÓ ®Æt trïng
khÝt lªn nhau hai vßng trßn cã b¸n kÝnh kh¸c nhau. HoÆc lµ Tù
nhiªn ®· sai: ®Æt c¸c kh«ng gian kh¸c nhau vµo trong nhau, bÊt
chÊp m©u thuÉn.

Vµ m©u thuÉn sinh ra do ®iÒu ®ã lµ ®éng lùc cña vËn ®éng, vËn ®éng
®Ó tho¸t ra khái m©u thuÉn.

Nh­ vËy, cã thÓ nãi vËt chÊt lµ tÊt c¶ c¸c thÓ kh«ng gian cã ®é
cong (tøc mèi quan hÖ néi t¹i) nµo ®ã.

Nh­ng tõ ®©u mµ sinh ra c¸c thÓ kh«ng gian nµy vµ lµm sao chóng cã
thÓ tån t¹i ®­îc?

Chóng ta h·y t­ëng t­îng r»ng tÊt c¶ biÕn mÊt hÕt: vËt chÊt, kh«ng
gian,... vµ nãi chung biÕn mÊt hÕt mäi sù kh¸c biÖt cã thÓ cã.

Khi ®ã cßn l¹i c¸i g×?

Ch¼ng cßn c¸i g× c¶!

Nh­ng ®ã chÝnh lµ c¸i duy nhÊt cßn l¹i!

Râ rµng c¸i duy nhÊt nµy lµ v« h¹n vµ ®ång nhÊt ë "kh¾p mäi n¬i".
NÕu kh«ng thÕ, sÏ vi ph¹m ®ßi hái cña chóng ta.

B©y giê chóng ta ®ßi hái mét ®iÒu tiÕp theo: ngay c¶ c¸i duy nhÊt
nµy còng biÕn mÊt nèt! SÏ cßn l¹i c¸i g× sau nã?

Kh«ng ph¶i vÊt v¶ l¾m, ta thÊy ngay r»ng c¸i ®Õn thay thÕ cho nã
l¹i lµ chÝnh nã! V× vËy ta h·y gäi c¸i ®ã lµ {\it kh«ng gian tuyÖt
®èi}.

Kh«ng gian tuyÖt ®èi cã thÓ biÕn mÊt vµo chÝnh nã, nãi kh¸c h¬n,
sù phñ nhËn nã dÉn tíi sù kh¼ng ®Þnh chÝnh nã. §iÒu ®ã cã nghÜa
lµ, c¸i kh«ng gian tuyÖt ®èi cña chóng ta cã thÓ tù tån t¹i mµ
kh«ng cÇn nhê ®Õn ai c¶. Nã lµ c¸i c¬ b¶n ®Çu tiªn.

Nã còng lµ "nguyªn nh©n tèi cao" n÷a. V× tr¸i víi mäi ý muèn cña
ai ®ã, nã vÉn chøa ®ùng sù kh¸c biÖt.

Qu¶ vËy, trong c¸i duy nhÊt ®ã kh«ng chøa c¸i g× c¶, vËy mµ vÉn
cã: c¸i {\it Kh«ng}! c¸i {\it Kh«ng} chøa trong c¸i {\it Cã}, c¸i
{\it Kh«ng} t¹o nªn c¸i {\it Cã}. Cã, nh­ng kh«ng lµ g× c¶!

{\VnTimeH ë} ®©y, sù phñ ®Þnh còng lµ sù kh¼ng ®Þnh, c¸i {\it
Kh«ng} còng lµ c¸i {\it Cã}, vµ ng­îc l¹i. M©u thuÉn néi t¹i cña
tr¹ng th¸i nµy lµ lín v« h¹n.

DiÔn ®¹t mét c¸ch to¸n häc h¬n ®iÒu ®ã: kh«ng gian tuyÖt ®èi cña
chóng ta cã ®é cong b»ng kh«ng. Trong kh«ng gian nµy tån t¹i c¸c
®iÓm mµ ®é cong cña chóng lµ v« cïng. Sù kh¸c biÖt nµy lín v« h¹n
vµ do ®ã m©u thuÉn sinh ra còng lín v« h¹n.

Tù nhiªn kh«ng muèn tån t¹i trong tr¹ng th¸i m©u thuÉn nh­ vËy. Nã
tù t×m c¸ch gi¶i quyÕt, vµ kÕt qu¶ lµ, thÕ giíi v× thÕ mµ ®­îc
sinh ra.

Nh­ vËy, l¹i mét lÇn n÷a, c¸i ch©n lý m¬ hå mµ ai còng quen thuéc
nÕu ®· qua c¸c tr­êng phæ th«ng lµ: "VËt chÊt kh«ng tù nhiªn sinh
ra (tøc lµ kh«ng sinh ra tõ h­ v«), kh«ng tù nhiªn mÊt ®i, lu«n
lu«n vËn ®éng vµ chuyÓn ho¸ tõ d¹ng nµy sang d¹ng kh¸c", nay cÇn
®­îc kh¼ng ®Þnh l¹i r»ng: "vËt chÊt" ®óng lµ sinh ra tõ "kh«ng cã
g×". Nh­ng kh«ng ph¶i v« cí mµ nh­ vËy. §éng lùc khiÕn nã sinh ra
còng lµ ®éng lùc khiÕn nã tån t¹i, biÕn ®æi, vµ vËn ®éng.

\begin{center}
{\bf BiÓu diÔn m©u thuÉn d­íi d¹ng ®Þnh l­îng:} \\ {\bf {\VnTimeH
Ph­¬ng tr×nh Nh©n qu¶}}
\end{center}

Mäi m©u thuÉn ®Òu ph¸t sinh bëi sù tån t¹i ®ång thêi cña hai kh¼ng
®Þnh lo¹i trõ nhau.

§iÒu ®ã ®­îc biÓu diÔn nh­ sau:
\[
  M = \left\{ \begin{array}{ll}
  A \neq B & \mbox{\it - Kh¼ng ®Þnh } K_1 \\
  A = B & \mbox{\it - Kh¼ng ®Þnh } K_2
  \end{array}\right.
\]
hoÆc
\[
  M = \left\{ \begin{array}{ll}
  A = A & \mbox{\it - Kh¼ng ®Þnh } K_1 \\
  A \neq A & \mbox{\it - Kh¼ng ®Þnh } K_2
  \end{array}\right.
\]

Râ rµng lµ m©u thuÉn sÏ cµng gay g¾t nÕu nh­ møc ®é phñ nhËn lÉn
nhau cña hai kh¼ng ®Þnh cµng lín. Nh­ng møc ®é phñ nhËn lÉn nhau
cña hai kh¼ng ®Þnh chØ cã thÓ ®­îc ®¸nh gi¸ b»ng møc ®é kh¸c biÖt
gi÷a hai kh¼ng ®Þnh ®ã. Bëi vËy ta cã thÓ biÓu diÔn m©u thuÉn mét
c¸ch t­îng tr­ng nh­ sau: $M = [K_1 - K_2]$.

M©u thuÉn ®­îc gi¶i quyÕt tøc lµ hiÖu $[K_1 - K_2]$ sÏ gi¶m tíi
kh«ng. §iÒu ®ã cã nghÜa lµ c¶ hai kh¼ng ®Þnh $K_1$ vµ $K_2$ ®Òu
ph¶i biÕn ®æi nh­ thÕ nµo ®ã ®Ó ®¹t tíi vµ kÕt thóc ë mét kh¼ng
®Þnh míi $K_3$.

Nh­ vËy, c¸c hiÖu $[K_1 - K_3]$ vµ $[K_2 - K_3]$ phô thuéc vµo
nh÷ng g×? Râ rµng c¸c hiÖu ®ã phô thuéc vµo søc b¶o toµn cña c¸c
kh¼ng ®Þnh $K_1$ vµ $K_2$. Søc b¶o toµn cña kh¼ng ®Þnh nµo cµng
lín th× sù kh¸c biÖt gi÷a nã vµ kh¼ng ®Þnh cuèi cïng cµng nhá.

VËy ®Õn l­ît m×nh, søc b¶o toµn cña mét kh¼ng ®Þnh nµo ®ã phô
thuéc vµo g×?

Cã hai yÕu tè. 1) Phô thuéc vµo m©u thuÉn néi t¹i cña kh¼ng ®Þnh
®ã, m©u thuÉn néi t¹i cµng lín søc b¶o toµn cña kh¼ng ®Þnh cµng
nhá. 2) Phô thuéc vµo m©u thuÉn míi, sinh ra do sù biÕn ®æi cña
kh¼ng ®Þnh. M©u thuÉn nµy cµng lín, sù biÕn ®æi cña kh¼ng ®Þnh
cµng bÞ c¶n trë vµ do ®ã søc b¶o toµn cña kh¼ng ®Þnh cµng lín.

Sù biÕn ®æi, mµ vËn ®éng lµ mét d¹ng cña nã, sinh ra do m©u thuÉn.
Nãi ®óng h¬n, vËn ®éng lµ biÓu hiÖn cña sù gi¶i quyÕt m©u thuÉn.

{\VnTimeH ë} trªn ta ®· nãi r»ng, m©u thuÉn cµng gay g¾t nÕu nh­
møc ®é kh¸c biÖt gi÷a hai kh¼ng ®Þnh lo¹i trõ nhau cµng lín. Chóng
ta sÏ tiÕp tôc lµm ®Çy ®ñ h¬n kÕt luËn ®Þnh l­îng nµy: m©u thuÉn
cµng gay g¾t th× nhu cÇu gi¶i tho¸t khái nã cµng bøc thiÕt vµ do
®ã sù vËn ®éng, biÕn ®æi cña tr¹ng th¸i, tøc lµ cña m©u thuÉn,
cµng quyÕt liÖt, nhanh chãng.

NÕu ta gäi sù quyÕt liÖt, hay ®é nhanh chãng biÕn ®æi cña m©u
thuÉn lµ $Q$, m©u thuÉn tr¹ng th¸i lµ $M$ th× cã thÓ biÓu diÔn
nguyªn lý trªn nh­ sau:
\[
Q \sim M \hspace{1cm} \mbox{vËy} \hspace{0.5cm} Q = K_{(M)}M.
\]

Ta h·y gäi ®ã lµ {\it ph­¬ng tr×nh nh©n qu¶}, trong ®ã $K_{(M)}$
lµ ph­¬ng tiÖn ®Ó gi¶i quyÕt m©u thuÉn.

Do c¸ch ®Æt vÊn ®Ò ë møc ®¬n gi¶n nhÊt nªn $K_{(M)}$ chØ cã thÓ lµ
mét hµm cña tr¹ng th¸i, tøc lµ hµm cña chÝnh m©u thuÉn. Thùc chÊt,
nã biÓu hiÖn møc ®é dÔ dµng cña sù tho¸t biÕn khái m©u thuÉn cña
tr¹ng th¸i.

NÕu m©u thuÉn ®­îc ®Æc tr­ng bëi c¸c ®¹i l­îng $..., x, y, z,...$
th× còng chÝnh nh÷ng ®¹i l­îng nµy sÏ lµ ph­¬ng tiÖn chuyÓn t¶i
m©u thuÉn, lµ c¸c bËc tù do mµ theo ®ã m©u thuÉn sÏ ®­îc gi¶i
quyÕt. Khi ®ã, møc ®é dÔ dµng cña sù tho¸t biÕn ph¶i ®­îc ®¸nh gi¸
nh­ lµ ®¹o hµm cña m©u thuÉn theo c¸c bËc tù do cña nã.

Gi¸ trÞ ®¹o hµm cña m©u thuÉn theo mét bËc tù do nµo ®ã cña nã
cµng lín th× kh¶ n¨ng "®¸nh h¬i" thÊy lèi tho¸t theo h­íng Êy cña
tr¹ng th¸i cµng lín, "l­îng m©u thuÉn" ®­îc gi¶i tho¸t theo bËc tù
do Êy cµng nhiÒu.

Nh­ vËy
\[
K_{(M)} \sim |M'(..., x, y, z, ...)|
\]
vµ ta cã
\begin{equation}
Q = a |M'(..., x, y, z, ...)| M(..., x, y, z, ...), \tag{I}
\label{NQ-pt}
\end{equation}
hÖ sè $a$ chØ cã thÓ sinh ra do sù lùa chän hÖ ®¬n vÞ cña c¸c ®¹i
l­îng.

Chóng ta ®· nãi r»ng, sù kh¸c biÖt lµ nguån gèc cña tÊt c¶. Nh­ng
sù kh¸c biÖt tù nã kh«ng cã nghÜa. C¸i gäi lµ "cã nghÜa" Êy chØ
sinh ra trong mèi quan hÖ trùc tiÕp, trong sù so s¸nh trùc tiÕp.
Tù nhiªn kh«ng thÓ c¶m biÕt ®­îc sù kh¸c biÖt qua "kho¶ng c¸ch".
Chóng ta thõa nhËn cã hai lo¹i kh¸c biÖt: chÊt vµ l­îng. §èi víi
Tù nhiªn, lo¹i kh¸c biÖt nµo lµ thùc sù tån t¹i?

Mét tr¹ng th¸i nµo ®ã nÕu cã m©u thuÉn néi t¹i th× nã ph¶i biÕn
®æi ®Ó ®¹t tíi tr¹ng th¸i kh«ng cã m©u thuÉn néi t¹i, hay ®óng
h¬n, ®¹t tíi tr¹ng th¸i cã m©u thuÉn néi t¹i nhá nhÊt cã thÓ ®­îc.

Qu¸ tr×nh ®ã lµ mét chiÒu, tr¶i qua liªn tôc tÊt c¶ c¸c gi¸ trÞ
cña m©u thuÉn, tõ gi¸ trÞ ban ®Çu ®Õn gi¸ trÞ cuèi cïng.

Nh­ vËy, chóng ta ®· cè g¾ng tù thuyÕt phôc r»ng vËn ®éng (biÕn
®æi) nhÊt thiÕt ph¶i cã nguyªn nh©n cña nã vµ tÝnh chÊt cña sù
biÕn ®æi tu©n theo ph­¬ng tr×nh nh©n qu¶. VËy th× sù kh«ng biÕn
®æi, tøc sù b¶o toµn, ph¶i lµ ®iÒu mÆc nhiªn, kh«ng cÇn cã nguyªn
nh©n ch¨ng? vµ cã thÓ nãi r»ng: bÊt kú tr¹ng th¸i nµo chØ cã hai
kh¶ n¨ng: hoÆc ®­îc b¶o toµn, hoÆc bÞ biÕn ®æi; hay ®óng h¬n, tÊt
c¶ ®Òu ®­îc b¶o toµn ({\it kh¼ng ®Þnh} $A$) nh­ng nÕu sù b¶o toµn
®ã g©y ra m©u thuÉn th× nã ph¶i nh­êng chç cho sù biÕn ®æi ®Ó
tho¸t khái m©u thuÉn vµ biÕn ®æi ®ã tu©n theo ph­¬ng tr×nh nh©n
qu¶.

NÕu luËn ®iÓm nµy ®óng th× c«ng viÖc cña chóng ta chØ lµ ë chç:
häc c¸ch hiÓu, ®¸nh gi¸ ®óng vµ ®Çy ®ñ m©u thuÉn cña tr¹ng th¸i,
m« t¶ nã theo ph­¬ng tr×nh nh©n qu¶, khi ®ã ta sÏ cã ®­îc quy luËt
cña bÊt kú sù biÕn ®æi nµo.

Nh­ng chØ cã thÕ th× ®· ®ñ ch­a, cho sù nhËn thøc ®Õn cïng cña
chóng ta vÒ Tù nhiªn, vÒ chÝnh b¶n th©n con ng­êi víi søc m¹nh t­
duy cña nã, ®Ó cã thÓ gi¶i thÝch ®iÒu kú diÖu, m·i m·i lµm ng¹c
nhiªn mäi thÕ hÖ: v× sao Tù nhiªn l¹i cã thÓ tù nhËn thøc ®­îc
chÝnh m×nh, th«ng qua s¶n phÈm cña nã: con ng­êi?!

\begin{center}
*

* \hspace{1cm} *
\end{center}

\begin{center}
{\bf {\VnTimeH ø}ng dông nguyªn lý nh©n qu¶ vµo mét vµi tr­êng
hîp}\\ {\bf cô thÓ, ®¬n gi¶n nhÊt}
\end{center}

Chóng ta h·y xem xÐt mét vµi thÝ dô ®¬n gi¶n nhÊt ®Ó minh ho¹ cho
nguyªn lý nh©n qu¶.

\noindent {\it 1. Ph­¬ng tr×nh truyÒn nhiÖt}

Gi¶ sö trong mét kho¶ng nµo ®ã cña kh«ng gian mét chiÒu ta cã ph©n
bè cña mét ®¹i l­îng $L$ nµo ®ã.

NÕu ph©n bè cã sù kh¸c biÖt néi t¹i - tøc lµ cã chøa m©u thuÉn néi
t¹i - th× ph©n bè sÏ biÕn ®æi ®Ó ®¹t tíi tr¹ng th¸i cã m©u thuÉn
néi t¹i nhá nhÊt. Sù biÕn ®æi ®ã tu©n theo ph­¬ng tr×nh nh©n qu¶
(\ref{NQ-pt}).

Ta ®­a vµo ®¹i l­îng $T$, nghÞch ®¶o cña $Q$, gäi lµ ®é tr× trÖ
cña sù gi¶i quyÕt m©u thuÉn. Nh­ vËy
\[
T = \frac 1{a |M'| M}.
\]

Tæng sè ®é tr× trÖ t¹o ra trong qu¸ tr×nh gi¶i quyÕt m©u thuÉn tõ
gi¸ trÞ ($M_0$) ®Õn gi¸ trÞ ($M_0 - \Delta M$) ta gäi lµ thêi
gian, ®­îc sinh ra do qu¸ tr×nh biÕn ®æi ®ã ($\Delta t$).
\begin{figure}[h]
\begin{center}
\includegraphics[width=1\columnwidth]{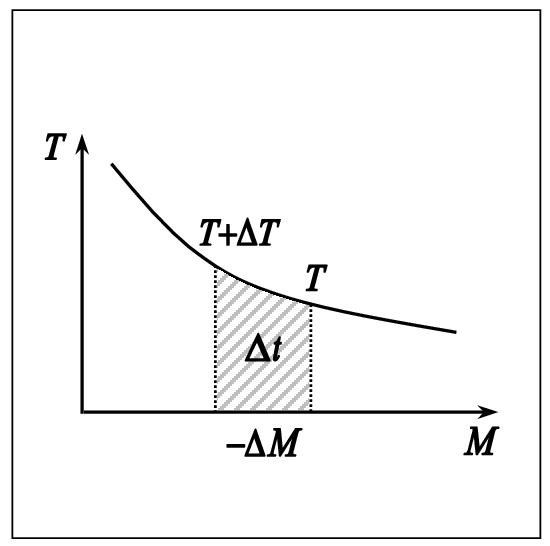}
\caption{\label{fig:h1}}
\end{center} 
\end{figure}

Theo ®Þnh nghÜa vµ tõ H×nh~\ref{fig:h1}, ta thÊy r»ng
\[
\Delta t \approx - \frac{2T + \Delta T}2 \ \Delta M,
\]
nh­ vËy
\[
\frac{\Delta M}{\Delta t} \approx - \frac 2{2T + \Delta T}.
\]
Ta cã
\[
\lim_{\Delta T \rightarrow 0, \Delta M \rightarrow 0, \Delta t
\rightarrow 0} \frac{\Delta M}{\Delta t} = \frac{dM}{dt} = - \frac
1 T = -a|M'|M.
\]
Ta thu ®­îc c¸ch diÔn ®¹t míi cña nguyªn lý nh©n qu¶:
\begin{equation}
\frac{dM}{dt} = - a |M'(..., x, y, z, ...)| M(..., x, y, z, ...).
\tag{II} \label{NQ-pt2}
\end{equation}

Nh­ vËy, nÕu chóng ta quy ­íc víi nhau ®Ó thêi gian trë thµnh mét
®¹i l­îng ®éc lËp, cßn m©u thuÉn l¹i lµ ®¹i l­îng phô thuéc nã th×
tèc ®é tho¸t biÕn theo thêi gian cña m©u thuÉn tû lÖ víi ®é lín
cña m©u thuÉn vµ ph­¬ng tiÖn gi¶i tho¸t nã.

Tr­êng hîp m©u thuÉn ®­îc ®Æc tr­ng bëi chÝnh nã, tøc lµ $M =
M_{(M)}$, ta cã
\[
M = M_0 e^{-a(t - t_0)}.
\]

Trë l¹i víi ph©n bè cña chóng ta. §Ó thuËn tiÖn, ta tr¶i ph©n bè
nµy theo trôc $x$ vµ lÊy mét ®iÓm nµo ®ã lµm gèc to¹ ®é.
\begin{figure}[h]
\begin{center}
\includegraphics[width=1\columnwidth]{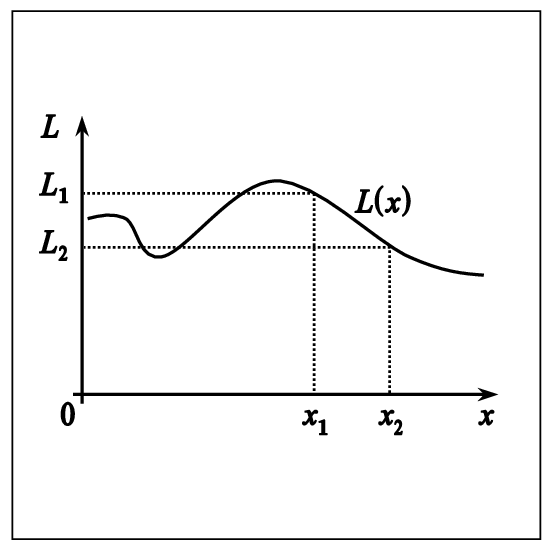}
\caption{\label{fig:h2}}
\end{center} 
\end{figure}

Bëi ph©n bè lµ cña mét ®¹i l­îng $L$ nµo ®ã cho nªn mäi gi¸ trÞ
cña nã t¹i c¸c ®iÓm cña kh«ng gian ph©n bè sÏ cã cïng thø nguyªn
(®ång nhÊt).

Sù kh¸c biÖt néi t¹i cña ph©n bè sÏ lµ sù kh¸c biÖt vÒ l­îng. T¹i
hai ®iÓm $x_1$ vµ $x_2$ ®¹i l­îng $L$ lÊy hai gi¸ trÞ $L_1$ vµ
$L_2$ t­¬ng øng. V× kh¸c biÖt vÒ l­îng nªn chØ cã mét c¸ch ®¸nh
gi¸ duy nhÊt: lÊy hiÖu $(L_2 - L_1)$.

Nh­ng hai ®iÓm $x_1$, $x_2$ chØ "c¶m thÊy sù kh¸c biÖt" cña nhau
trong mèi quan hÖ trùc tiÕp, m©u thuÉn xuÊt hiÖn hay kh«ng chØ
trong mèi quan hÖ trùc tiÕp ®ã: t¹i ranh giíi cña hai ®iÓm l©n cËn
$x_1$, $x_2$ ®¹i l­îng $L$ ®ång thêi lÊy hai gi¸ trÞ $L_1$, $L_2$,
hai kh¼ng ®Þnh nµy phñ nhËn lÉn nhau vµ ®é lín cña m©u thuÉn phô
thuéc vµo hiÖu $(L_2 - L_1)$. Bëi vËy, ®Ó cho hiÖu $(L_2 - L_1)$
lµ s¶n phÈm cña mèi quan hÖ trùc tiÕp gi÷a hai ®iÓm $x_1$, $x_2$
th× ta ph¶i cho, ch¼ng h¹n, ®iÓm $x_2$ tiÕn gÇn v« h¹n tíi ®iÓm
$x_1$ (nh­ng kh«ng trïng víi nã).

Khi ®ã, m©u thuÉn néi t¹i t¹i khu vùc ®iÓm $x_1$ sÏ ®­îc ®¸nh gi¸
nh­ lµ giíi h¹n cña tØ sè $\frac{L_2 - L_1}{x_2 - x_1}$ khi $x_2
\rightarrow x_1$, tøc lµ b»ng gi¸ trÞ ®¹o hµm cña ®¹i l­îng $L$
theo kh«ng gian ph©n bè t¹i ®iÓm $x_1$.

Tõ nh÷ng ®iÒu ®· tr×nh bµy, ta cã $M = \frac{dL}{dx} =
\frac{\partial L}{\partial x}$. Thay gi¸ trÞ cña $M$ vµo ph­¬ng
tr×nh nh©n qu¶ (\ref{NQ-pt2}):
\begin{equation}
\frac{\partial}{\partial t}\frac{\partial L}{\partial x} = -a
\frac{\partial L}{\partial x}. \label{pt1}
\end{equation}

M©u thuÉn néi t¹i t¹i mçi ®iÓm ®­îc gi¶i quyÕt theo ph­¬ng tr×nh
(\ref{pt1}). §iÒu ®ã lµm ph©n bè bÞ biÕn ®æi. Ta h·y t×m quy luËt
cña sù biÕn ®æi nµy.

M©u thuÉn néi t¹i t¹i khu vùc ®iÓm $x$ lµ $M_{x, t} =
\left.\frac{\partial L}{\partial x}\right|_{x, t}$.

Sau kho¶ng thêi gian $\Delta t$ m©u thuÉn nµy gi¶m xuèng ®Õn gi¸
trÞ $M_{x, t+\Delta t} = \left.\frac{\partial L}{\partial
x}\right|_{x, t+\Delta t}$.
\begin{figure}[h]
\begin{center}
\includegraphics[width=1\columnwidth]{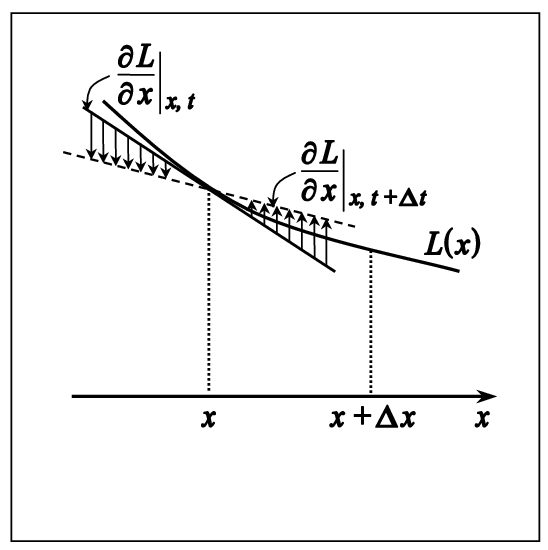}
\caption{\label{fig:h3}}
\end{center} 
\end{figure}

Nh­ vËy, h×nh nh­ lµ sù biÕn ®æi nµy ®· dån Ðp mét l­îng nµo ®ã
gi¸ trÞ cña ®¹i l­îng $L$ tõ c¸c ®iÓm cã gi¸ trÞ cao h¬n sang
nh÷ng ®iÓm cã gi¸ trÞ thÊp h¬n, t¹o nªn mét "dßng ch¶y" gi¸ trÞ
cña ®¹i l­îng $L$ qua ®iÓm $x$. Trong vÝ dô cña chóng ta "dßng
ch¶y" ®ã ch¶y tõ nh÷ng ®iÓm bªn tr¸i h¬n ®iÓm $x$ sang nh÷ng ®iÓm
bªn ph¶i h¬n nã.

Râ rµng, ®é lín cña "dßng ch¶y", tøc lµ l­îng c¸c gi¸ trÞ $L$ ch¶y
qua ®iÓm $x$ trong kho¶ng thêi gian $\Delta t$ lµ
\[
{\cal J}_x = \left.\frac{\partial}{\partial t}\frac{\partial
L}{\partial x}\right|_x \Delta t = - a \left.\frac{\partial
L}{\partial x}\right|_x \Delta t.
\]
T­¬ng tù nh­ vËy, t¹i ®iÓm $x + \Delta x$, ta cã
\[
{\cal J}_{x+\Delta x} = - a \left.\frac{\partial L}{\partial
x}\right|_{x +\Delta x} \Delta t.
\]

Trong vÝ dô trªn, dßng ${\cal J}_x$ lµm gi¸ trÞ ®¹i l­îng $L$ t¹i
c¸c ®iÓm trong kho¶ng $\Delta x$ t¨ng lªn, cßn dßng ${\cal J}_{x
+\Delta x}$ th× lµm chóng gi¶m xuèng. KÕt qu¶ lµ sè gia $\Delta L$
mµ kho¶ng $\Delta x$ nhËn ®­îc lµ
\begin{eqnarray*}
\left.\Delta L\right|_{\Delta t} &=& a \Delta t \left(
\left.\frac{\partial L}{\partial x}\right|_{x + \Delta x} -
\left.\frac{\partial L}{\partial x}\right|_x \right) \\ &=& a
\Delta t \left.\frac{\partial^2 L}{\partial x^2}\right|_{x\leq \xi
\leq x +\Delta x} \Delta x.
\end{eqnarray*}
MËt ®é trung b×nh gi¸ trÞ $\overline{\Delta L}$ t¹i mçi ®iÓm trong
kho¶ng $\Delta x$ sÏ lµ
\[
\left.\overline{\Delta L}\right|_{\Delta t} \cong \frac{a \Delta t
\left.\frac{\partial^2 L}{\partial x^2}\right|_\xi \Delta
x}{\Delta x}.
\]
Gi¸ trÞ chÝnh x¸c ®¹t ®­îc ë giíi h¹n
\[
\left. \Delta L \right|_{x, \Delta t} = \lim_{\Delta x \rightarrow
0} \left.\overline{\Delta L}\right|_{\Delta t} = a \Delta t
\left.\frac{\partial^2 L}{\partial x^2}\right|_x.
\]
Nh­ vËy,
\[
\lim_{\Delta t \rightarrow 0} \left.\frac{\Delta L}{\Delta
t}\right|_x = a \frac{\partial^2 L}{\partial x^2},
\]
hay
\begin{equation}
\frac{\partial L}{\partial t} = a \frac{\partial^2 L}{\partial
x^2}. \label{pt2}
\end{equation}

Tèc ®é biÕn ®æi theo thêi gian cña ®¹i l­îng $L$ t¹i l©n cËn cña
bÊt kú ®iÓm nµo cña ph©n bè tØ lÖ víi ®¹o hµm bËc hai theo kh«ng
gian ph©n bè cña ®¹i l­îng nµy t¹i chÝnh ®iÓm Êy.

Nh­ng nh­ ta ®· biÕt, ph­¬ng tr×nh (\ref{pt2}) chÝnh lµ ph­¬ng
tr×nh truyÒn nhiÖt mµ vËt lý häc ®· t×m ra.

MÆt kh¸c, hÖ qu¶ cña c¸ch lËp luËn trªn cho ta sù b¶o toµn gi¸ trÞ
cña ®¹i l­îng $L$ trong toµn bé ph©n bè, dï r»ng gi¸ trÞ cña ®¹i
l­îng nµy t¹i mçi ®iÓm riªng biÖt cã thÓ biÕn ®æi, hÔ gi¸ trÞ t¹i
®iÓm nµy gi¶m ®i mét l­îng nµo ®ã th× gi¸ trÞ cña ®iÓm l©n cËn
t¨ng lªn ®óng mét l­îng nh­ vËy. NÕu kh«ng gian ph©n bè lµ v« h¹n,
th× cïng víi sù t¨ng lªn cña thêi gian, gi¸ trÞ trung b×nh cña
ph©n bè dÇn gi¶m tíi kh«ng.

\noindent {\it 2. Con quay håi chuyÓn}

XÐt mét vÝ dô kh¸c. Chóng ta cã hai con quay víi m«men ®éng l­îng
t­¬ng øng lµ $k_1 \vec{\omega}_1$ vµ $k_2 \vec{\omega}_2$. Sù tån
t¹i cña hai con quay víi sù b¶o toµn cña c¸c m«men ®éng l­îng cña
chóng, xÐt theo gãc ®é vÜ m«, lµ c¸c kh¼ng ®Þnh $K_1$ vµ $K_2$.

\begin{figure}[h]
\begin{center}
\includegraphics[width=1\columnwidth]{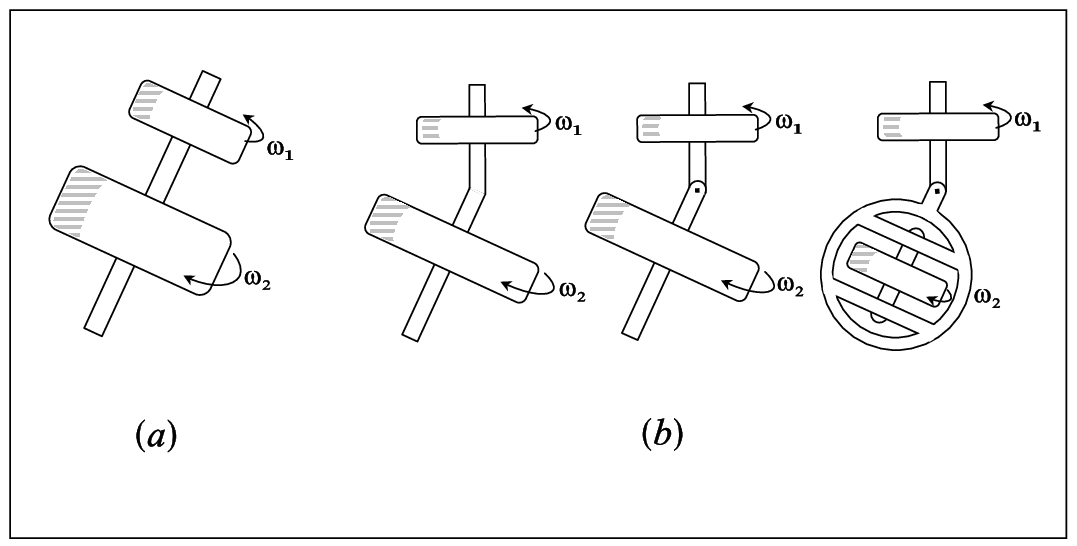}
\caption{\label{fig:h4}}
\end{center} 
\end{figure}

Sù b¶o toµn cña c¸c vector m«men ®éng l­îng cã thÓ coi lµ sù b¶o
toµn cña hai yÕu tè: b¶o toµn ph­¬ng vµ b¶o toµn ®é lín. NÕu chóng
®­îc g¾n víi nhau theo H×nh~\ref{fig:h4}.a th× sù b¶o toµn ph­¬ng
cña chóng kh«ng bÞ x©m ph¹m, nh­ng sù b¶o toµn ®é lín cña mét
vector sÏ bÞ sù b¶o toµn cña vector kia x©m ph¹m. M©u thuÉn sÏ
cµng gay g¾t nÕu sù kh¸c biÖt gi÷a hai ®é lín cña c¸c vector nµy
cµng lín. KÕt qu¶ lµ hÖ thèng nµy ph¶i biÕn ®æi thÕ nµo ®ã ®Ó c¶
hÖ sÏ cã mét vector m«men ®éng l­îng duy nhÊt $k_3
\vec{\omega}_3$. Trong tr­êng hîp hai con quay ®­îc g¾n víi nhau
nh­ trong c¸c H×nh~\ref{fig:h4}.b m©u thuÉn tr¹ng th¸i sÏ phøc t¹p
h¬n. Kh«ng chØ sù b¶o toµn vÒ ®é lín mµ c¶ sù b¶o toµn vÒ ph­¬ng
®Òu bÞ x©m ph¹m. C¸ch thøc gi¶i quyÕt m©u thuÉn cña tr¹ng th¸i phô
thuéc vµo kÕt cÊu cña khíp nèi.

\begin{figure}[h]
\begin{center}
\includegraphics[width=1\columnwidth]{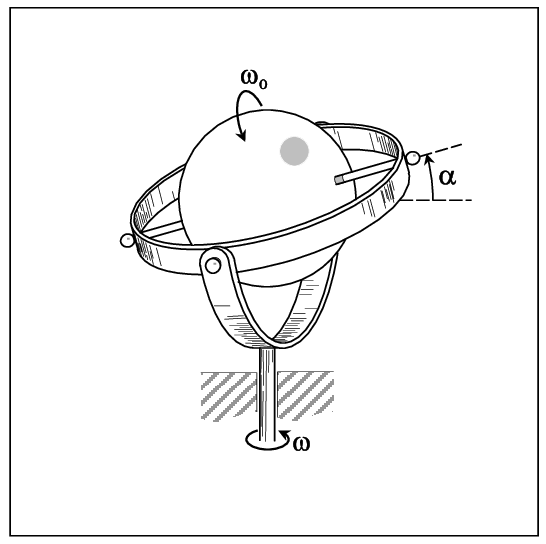}
\caption{\label{fig:h5}}
\end{center}
\end{figure}

Chóng ta xÐt tr­êng hîp thø ba, ë ®ã c¸c yÕu tè träng lùc vµ ly
t©m (cã trong tr­êng hîp thø hai) cã thÓ coi nh­ kh«ng ®¸ng kÓ
(H×nh~\ref{fig:h5}).

§Ó ®¬n gi¶n, chóng ta cho r»ng ®éng c¬ duy tr× vËn tèc gãc
$\omega$ cña hÖ kh«ng ®æi. Nh­ vËy chóng ta sÏ chØ quan t©m ®Õn
m©u thuÉn sinh ra do sù b¶o toµn ph­¬ng cña $k \vec{\omega}_0$ bÞ
x©m ph¹m.

Kh¼ng ®Þnh $K_1 \equiv$ sù b¶o toµn cña $k \vec{\omega}_0$, nãi
r»ng: tèc ®é biÕn thiªn cña ph­¬ng vector $k \vec{\omega}_0$ b»ng
kh«ng. Nh­ng kh¼ng ®Þnh $K_2 \equiv$ sù b¶o toµn cña $\omega$, nãi
r»ng: kh«ng, ph­¬ng cña $k \vec{\omega}_0$ ph¶i biÕn ®æi víi vËn
tèc gãc $\omega \cos\alpha$.

Nh­ vËy, ë møc ®é vÜ m«, hiÖu $[K_1 - K_2] = \omega \cos\alpha$ lµ
nguån gèc cña m©u thuÉn vµ m©u thuÉn ®ã tØ lÖ víi hiÖu nµy,
\[
M \sim \omega\cos\alpha; \hspace{1cm} M = k\omega_0
\omega\cos\alpha,
\]
hÖ sè tØ lÖ $k\omega_0$ ®­îc ®­a vµo (vÉn trªn quan ®iÓm vÜ m«)
dùa trªn lý lÏ: nÕu $\omega_0$ b»ng kh«ng th× ph­¬ng cña vector $k
\vec{\omega}_0$ kh«ng tån t¹i mét c¸ch x¸c ®Þnh vµ do ®ã vÊn ®Ò
m©u thuÉn ph¸t sinh do sù b¶o toµn ph­¬ng cña nã kh«ng ®­îc ®Æt
ra.

Thay gi¸ trÞ cña $M$ vµo ph­¬ng tr×nh nh©n qu¶ (\ref{NQ-pt2}), ta
®­îc
\[
\frac{\partial M}{\partial t} = -a k^2 \omega_0^2 \omega^2
\sin\alpha \cos\alpha.
\]
Tõ ph­¬ng tr×nh ta thÊy r»ng, nÕu $\alpha =0$ th× tèc ®é tho¸t
biÕn cña m©u thuÉn tr¹ng th¸i b»ng kh«ng.

LÊy ®¹o hµm cña m©u thuÉn theo thêi gian,
\begin{equation}
\frac{\partial \alpha}{\partial t} = a k \omega_0 \omega
\cos\alpha, \hspace{1cm} (\alpha \neq 0). \label{pt3}
\end{equation}

Sù biÕn ®æi cña $\alpha$ g©y ra m©u thuÉn míi, m©u thuÉn nµy tû lÖ
víi gi¸ trÞ cña $\frac{\partial \alpha}{\partial t}$, do ®ã sÏ
kh«ng cã sù b¶o toµn vËn ®éng theo yÕu tè $\alpha$, vµ nh­ vËy tèc
®é tho¸t biÕn trong c«ng thøc (\ref{pt3}) chÝnh lµ vËn tèc tøc
thêi cña trôc mÆt ph¼ng quay theo yÕu tè $\alpha$.

Thêi gian ®Ó gãc gi÷a trôc mÆt ph¼ng quay (tøc ph­¬ng cña vector
$k \vec{\omega}_0$) vµ ph­¬ng n»m ngang biÕn ®æi tõ gi¸ trÞ ($+0$)
®Õn gi¸ trÞ ($\alpha$) sÏ lµ
\[
t = \frac 1{2ak\omega_0 \omega} \left.\ln \frac{1 + \sin\alpha}{1
- \sin\alpha}\right|_{+0}^\alpha.
\]

\noindent {\it 3. VÒ b¶n chÊt cña c¸i gäi lµ "tr­êng"}

VËt lý häc hiÖn nay cho r»ng thÕ giíi ®­îc cÊu t¹o bëi c¸c h¹t c¬
b¶n. VËy th× mét vÊn ®Ò ®­îc ®Æt ra lµ: c¸c h¹t c¬ b¶n ®ã ph¶i cã
cÊu tróc néi t¹i thÕ nµo ®Ó cã thÓ b¶o toµn ®­îc ®èi víi nguyªn lý
nh©n qu¶?

Cã thÓ gi¶i quyÕt vÊn ®Ò nµy mét c¸ch thuÇn tuý lý thuyÕt ®­îc
kh«ng?

Cho r»ng cã mét kh«ng gian h÷u h¹n $[A]$ cã cÊu tróc néi t¹i tho¶
m·n sù bÊt biÕn ®èi víi nguyªn lý nh©n qu¶.

Kh«ng gian nµy n»m trong kh«ng gian tuyÖt ®èi [O] cña chóng ta.
T¹i n¬i ranh giíi cña hai kh«ng gian xuÊt hiÖn m©u thuÉn do sù
kh¸c biÖt gi÷a hai kh«ng gian ®ã g©y ra.

Bëi c¶ hai kh«ng gian ®Òu tù b¶o toµn nªn m©u thuÉn ®ã chØ cã thÓ
®­îc gi¶i quyÕt b»ng c¸ch h×nh thµnh mét vïng ®Öm (tøc tr­êng),
nhê ®ã sù kh¸c biÖt trë lªn dÞu h¬n, ®iÒu hoµ h¬n. CÊu tróc cña
vïng ®Öm ph¶i thÕ nµo ®ã ®Ó møc ®é ®iÒu hoµ ®¹t tíi gi¸ trÞ lín
nhÊt, tøc lµ m©u thuÉn néi t¹i t¹i mçi ®iÓm cña tr­êng cã gi¸ trÞ
nhá nhÊt cã thÓ ®­îc.

Mét ®iÒu râ rµng lµ cµng xa t©m cña kh«ng gian $[A]$, tÝnh chÊt
$[A]$ cµng gi¶m ®i. Nãi c¸ch kh¸c, vïng ®Öm (tr­êng) bao quanh
kh«ng gian $[A]$ còng cã tÝnh chÊt $[A]$ nh­ng tÝnh chÊt nµy lµ
hµm cña $r$, tøc kho¶ng c¸ch tõ ®iÓm ®ang xÐt cña tr­êng tíi t©m
cña kh«ng gian $[A]$.

Tõ nh÷ng ®iÒu ®· tr×nh bµy vµ nÕu ký hiÖu vïng ®Öm lµ $T$, ta cã
\[
T_{[A]} = g(r) \frac{[A]}{r},
\]
$g(r)$ lµ mét hµm ch­a biÕt, nã ®Æc tr­ng cho sù hµi hoµ néi t¹i
cña tr­êng.
\begin{figure}[h]
\begin{center}
\includegraphics[width=1\columnwidth]{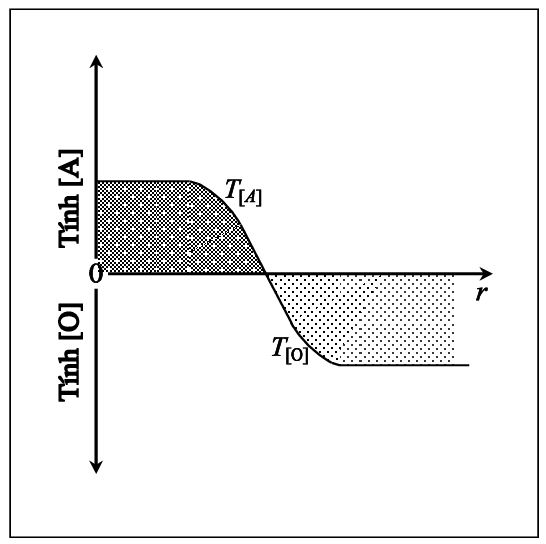}
\caption{\label{fig:h6}}
\end{center}
\end{figure}

NÕu trong vïng tr­êng $T_{[A]}$ cã mét kh«ng gian $[B]$ vµ kh«ng
gian nµy kh«ng lµm nhiÔu lo¹n ®¸ng kÓ tr­êng $T_{[A]}$, th× khi ®ã
sù kh¸c biÖt gi÷a $[B]$ vµ $T_{[A]}$ sÏ buéc $[B]$ ph¶i vËn ®éng
trong tr­êng ®Ó ®¹t tíi vÞ trÝ mµ ë ®ã kh¸c biÖt gi÷a $[B]$ vµ
$T_{[A]}$ cã trÞ sè nhá nhÊt (ë ®©y chóng ta ®· cho r»ng kh«ng
gian $[B]$ còng cã kh¶ n¨ng tù b¶o toµn). M©u thuÉn tr¹ng th¸i ®ã
tû lÖ víi hiÖu $\left[ [B] - T_{[A]}\right]$.

NÕu t×m ®­îc hÖ sè $c$ dïng ®Ó "dÞch ng«n ng÷" cña tÝnh chÊt $[B]$
sang "ng«n ng÷" cña tÝnh chÊt $[A]$ th× m©u thuÉn cã thÓ ®­îc diÔn
®¹t nh­ sau:
\[
M = f\left( c[B] - g(r) \frac{[A]}{r}\right), \hspace{1cm} f -
\mbox{hÖ sè tû lÖ.}
\]

Quy luËt vËn ®éng cña kh«ng gian $[B]$ trong tr­êng $T_{[A]}$ sÏ
®­îc t×m ra qua ph­¬ng tr×nh nh©n qu¶ (\ref{NQ-pt2}):
\begin{eqnarray*}
\frac{\partial M}{\partial t} &=& -a |M'|M  \\ &=& -a[A]f^2 \left|
\frac{g(r) - rg'(r)}{r^2}\right| \left( c[B] - [A]
\frac{g(r)}r\right).
\end{eqnarray*}
{\VnTimeH ë} ®©y, ®¹i l­îng chuyÓn t¶i (bËc tù do) cña m©u thuÉn
lµ $r$.

Bëi sù vËn ®éng cña kh«ng gian $[B]$ ph¶i x¶y ra ®ång thêi theo
tÊt c¶ c¸c ph­¬ng cã thµnh phÇn h­íng t©m, do ®ã vËn tèc tho¸t
biÕn tæng hîp cña tr¹ng th¸i - tøc lµ vËn tèc tæng hîp cña kh«ng
gian $[B]$ trong tr­êng $T_{[A]}$ sÏ ®­îc ®¸nh gi¸ nh­ lµ tÝch
ph©n cña tèc ®é tho¸t biÕn theo mäi ph­¬ng cã thµnh phÇn h­íng
t©m.
\begin{eqnarray*}
\frac{\partial M}{\partial t} \!\! &=&\!\! -a[A] 4\pi f^2\!\!\!\!
\int_0^{\frac{\pi}{2}}\!\! \left| \frac{g(r) \!-\! r
g'\!(r)}{r^2}\right| \!\! \left( \! c[B] \!-\! [A] \frac{g(r)}r\!
\right)\! \cos^2\!\!\varphi d\varphi \\ &=& -a\pi^2 f^2 [A] \left|
\frac{g(r) \!-\! r g'\!(r)}{r^2}\right| \!\!\left( \! c[B] \!-\!
[A] \frac{g(r)}r\! \right).
\end{eqnarray*}
Khai triÓn vÕ tr¸i ta ®­îc
\begin{eqnarray*}
f [A] \frac{g(r)}{r^2} \frac{\partial r}{\partial t} &=& -a\pi^2
f^2 [A] \left| \frac{g(r) \!-\! r g'\!(r)}{r^2}\right| \!\!\left(
\! c[B] \!-\! [A] \frac{g(r)}r \! \right) \\ \frac{\partial
r}{\partial t} &=& -a f \pi^2 \frac{\left| g(r) - r g'(r)
\right|}{g(r)} \left( c[B] - [A] \frac{g(r)}r\right).
\end{eqnarray*}


NÕu chøng tá ®­îc r»ng sù biÕn ®æi cña $r$ còng nh­ sù b¶o toµn
cña $\frac{\partial r}{\partial t}$ g©y ra m©u thuÉn míi tû lÖ víi
chÝnh $\frac{\partial r}{\partial t}$, th× tèc ®é tho¸t biÕn võa
thu ®­îc còng chÝnh lµ vËn tèc tøc thêi cña $[B]$ trong tr­êng
$T_{[A]}$.

\end{multicols}

\hspace{-0.4cm}\rule{.1mm}{2mm}\rule{18cm}{.1mm}\rule{.1mm}{2mm}

\end{document}